\documentclass[aps,prb,twocolumn,showpacs,preprintnumbers,amsmath,amssymb,superscriptaddress]{revtex4-1}
\usepackage{graphicx}
\usepackage{bm}

\begin{document}

\preprint{PREPRINT (\today)}

\title{Superconducting parameters of BaPt$_{4-x}$Au$_x$Ge$_{12}$ filled
skutterudite}

\author{A.\,Maisuradze}
\email{alexander.maisuradze@psi.ch} \affiliation{Laboratory for Muon Spin
Spectroscopy, Paul Scherrer Institut, CH-5232 Villigen PSI, Switzerland}
\affiliation{Physik-Institut der Universit\"at Z\"urich, Winterthurerstrasse
190, CH-8057 Z\"urich, Switzerland}
\author{R.\,Gumeniuk}
\affiliation{Max-Planck-Institut f\"ur Chemische Physik fester
Stoffe, N\"othnitzer Str.\ 40, 01187 Dresden, Germany}
\author{W.\,Schnelle}
\affiliation{Max-Planck-Institut f\"ur Chemische Physik fester
Stoffe, N\"othnitzer Str.\ 40, 01187 Dresden, Germany}
\author{M.\,Nicklas}
\affiliation{Max-Planck-Institut f\"ur Chemische Physik fester
Stoffe, N\"othnitzer Str.\ 40, 01187 Dresden, Germany}
\author{C.\,Baines}
\affiliation{Laboratory for Muon Spin Spectroscopy, Paul Scherrer
Institut, CH-5232 Villigen PSI, Switzerland}
\author{R.\,Khasanov}
\affiliation{Laboratory for Muon Spin Spectroscopy, Paul Scherrer
Institut, CH-5232 Villigen PSI, Switzerland}
\author{A.\,Amato}
\affiliation{Laboratory for Muon Spin Spectroscopy, Paul Scherrer
Institut, CH-5232 Villigen PSI, Switzerland}
\author{A.\,Leithe-Jasper}
\affiliation{Max-Planck-Institut f\"ur Chemische Physik fester
Stoffe, N\"othnitzer Str.\ 40, 01187 Dresden, Germany}

\begin{abstract}
We report on a study of the superconducting properties for a series of
polycrystalline BaPt$_{4-x}$Au$_x$Ge$_{12}$ filled skutterudite compounds for
$x = 0$, 0.5, 0.75, and 1. Muon spin rotation ($\mu$SR) spectroscopy as well
as magnetization, specific heat, and electrical resistivity measurements were
performed. The magnetic penetration depth $\lambda$, the coherence length
$\xi$, and the Ginzburg-Landau parameter $\kappa$ are evaluated. The
temperature dependence of the superfluid density is well described by an
$s$-wave superconducting gap and this classical scenario is supported by the
field-independent $\lambda$. The gap-to-$T_c$ ratio $\Delta_0/k_BT_c$
increases with the Au content from 1.70 for $x = 0$ to 2.1(1) for $x = 1$. By
combining $\mu$SR, magnetization, and specific heat data, we find that
BaPt$_{4-x}$Au$_x$Ge$_{12}$ compounds are in between the dirty and clean
limits with mean free paths of the carriers $l \sim \xi$. Interestingly,
resistivity data for BaPt$_4$Ge$_{12}$ indicate a much higher upper critical
field, which is probably due to defects or impurities close to the surface of
the crystallites.
\end{abstract}


\maketitle

\section{Introduction}

Filled skutterudite compounds $MT_4X_{12}$ with a framework formed from $T$
(Fe, Ru, Os) and $X$ (P, As, Sb) atoms and ``filled'' with $M$ atoms
(rare-earth, alkaline-earth, or alkali metals) came in focus of recent
research activities due to a number of unconventional
phenomena.\cite{Sales03rev,Aoki05rev,Maple05rev,Sato07rev,Uher2001,Jeitschko1977,Jasper2004}
In their cubic crystal structure, the filler cations $M$ reside in
icosahedral cages formed by tilted $TX_6$ octahedra. The pronounced
vibrational amplitudes of the $M$ atoms have been linked to dynamic
scattering mechanisms for heat-carrying acoustic phonons resulting in a
reduced lattice thermal conductivity, a prerequisite for thermoelectric
applications.\cite{Nolas1999} Several filled skutterudites display
superconductivity and as well show a broad variety of other interesting
phenomena.\cite{Meisner8185,Uchiumi99,EDBauer02,Shirotani03,Shirotani05}

Recently, a new family of filled skutterudites based on a different framework
of platinum and germanium with the chemical formula $M$Pt$_4$Ge$_{12}$ has
been discovered.\cite{Bauer07,Gumeniuk08} Several of these compounds are
superconducting (SC). The compositions with $M$ = Sr and
Ba\cite{Bauer07,Gumeniuk08} have SC transition temperatures $T_c$ around 5\,K
and the later reported ThPt$_4$Ge$_{12}$ is SC below
4.62\,K.\cite{Kaczorowski08,Bauer08Th} Due to a peak in the electronic
density of states (DOS) at the Fermi energy ($E_F$), LaPt$_4$Ge$_{12}$ has a
significantly higher $T_c$ of 8.3\,K.\cite{Gumeniuk08} Interestingly,
PrPt$_4$Ge$_{12}$, with trivalent Pr in a non-magnetic crystal field ground
state, is also SC with an only slightly lower $T_c$ of 7.9\,K. Its SC
properties\cite{Maisuradze09PRL,Maisuradze10PRB} show some similarities with
the heavy-fermion superconductivity of
PrOs$_4$Sb$_{12}$.\cite{EDBauer02,Sugaware02R,Aoki03,Maple07both} Most
remarkably, an unconventional SC order parameter with point nodes and a
rather similar gap-to-$T_c$ ratio has been observed. Moreover, signatures of
time-reversal-symmetry breaking were found in PrPt$_4$Ge$_{12}$ by zero-field
$\mu$SR.\cite{Maisuradze10PRB}

For LaPt$_4$Ge$_{12}$, SrPt$_4$Ge$_{12}$, and BaPt$_4$Ge$_{12}$, NMR and NQR
studies suggested an $s$-wave BCS SC state with $\Delta_0/k_BT_c \approx
1.60$.\cite{Kanetake09,Toda08a,Magishi09} Theoretical and experimental studies
of the electronic structure of this class of skutterudites consistently show
rather deep-lying Pt 5$d$ states which only partially form covalent bands with
Ge 4$p$ electrons.\cite{Gumeniuk2010} In turn, the electronic states at $E_F$
that are relevant for the SC behavior, can be firmly assigned to originate
predominantly from Ge 4$p$ electrons.\cite{Rosner2009} Different to
$M$Pt$_4$Ge$_{12}$ ($M$ = La, Pr) a pronounced peak in the DOS is located
little above $E_F$ for SrPt$_4$Ge$_{12}$ and
BaPt$_4$Ge$_{12}$.\cite{Gumeniuk08AuOpt} Here, the low-lying Pt states open up
the chance to influence the Fermi level in a rigid-band like manner by a
suitable substitution of Pt. By electron doping, the DOS and thus the SC $T_c$
of BaPt$_4$Ge$_{12}$ could be systematically influenced through substitution
of Pt by Au. The $T_c$ in the series BaPt$_{4-x}$Au$_x$Ge$_{12}$ could be
optimized to 7.0\,K for $x = 1$.\cite{Gumeniuk08AuOpt} For a doping with more
than 1.0 extra electrons per formula unit a decrease of $T_c$ is
expected.\cite{Gumeniuk08AuOpt} Actually, the partial substitution of Au for
Pt in LaPt$_4$Ge$_{12}$ leads to a continuous decrease of
$T_c$.\cite{LingweiLi10a}

Here, we report on a study of BaPt$_{4-x}$Au$_x$Ge$_{12}$ compounds ($x = 0$,
0.5, 0.75, and 1) by means of transverse field (TF) muon-spin rotation
($\mu$SR) spectroscopy and macroscopic magnetization, specific heat, and
electrical resistivity measurements. High-quality $\mu$SR spectra of
well-ordered SC flux-line lattices allowed us to use the exact solution of the
Ginzburg-Landau (GL) equations for the analysis. The superfluid density
($\rho_s$) was found to saturate exponentially in the low-temperature limit,
suggesting a SC gap without nodes, in agreement with a NMR
study.\cite{Magishi09} The temperature dependence of $\rho_s$ is well
described by the $s$-wave BCS function with the gap-to-$T_c$ ratios
($\Delta_0/k_BT_c$) of 1.70, 2.07, 2.15, and 2.02 for $x = 0$, 0.5, 0.75, and
1, respectively. This clear increase of $\Delta_0/k_BT_c$ (viz.\
electron-phonon coupling) with $x$ is in agreement with the results of our
previous study.\cite{Gumeniuk08AuOpt} The BCS character of the
superconductivity is further supported by the field-independent magnetic
penetration depth ($\lambda$). By combining $\mu$SR, magnetization and
specific heat data we find that BaPt$_{4-x}$Au$_x$Ge$_{12}$ compounds are in
between the dirty and clean limits with mean free path of the carriers $l \sim
\xi$. In electrical resistivity measurements on BaPt$_4$Ge$_{12}$ we observe
superconductivity for fields much higher than the (bulk) upper critical field.
This discrepancy appears especially for $x = 0$ samples and its origin is
discussed in terms of the presence of chemical or crystallographic defects
close to or on the surface of the crystallites.

The paper is organized as follows: in Section II we give some experimental
details, Sec.\ III describes the method of analysis of our $\mu$SR data, then
we present and discuss the results from the $\mu$SR as well as from the
macroscopic methods. Our conclusions are given in Sec.\ IV. In the Appendix we
describe the details of our calculations and give the relevant GL definitions.


\section{Experimental details}

Polycrystalline samples of BaPt$_{4-x}$Au$_x$Ge$_{12}$ with bulk $T_c$ values
of 4.9(1), 5.3(1), 6.25(5), and 6.95(5)\,K for $x = 0$, 0.5, 0.75, and 1,
respectively, were prepared as described in Refs.\ \onlinecite{Gumeniuk08}
and \onlinecite{Gumeniuk08AuOpt}. The SC transition temperatures $T_c$ were
determined from the onset of the Meissner flux expulsion (field cooling;
tangent to the steepest slope and extrapolation to $\chi = 0$) in magnetic
susceptibility data measured in a nominal field of 2\,mT (MPMS-XL7, Quantum
Design).

The transverse field (TF) $\mu$SR experiments were performed at the $\pi$M3
and $\mu$E1 beam lines at the Paul Scherrer Institute (Villigen Switzerland)
at the GPS, the LTF, and the GPD spectrometers. Each sample
used for the $\mu$SR study has an ellipsoid-like shape of a droplet with
dimensions: $\simeq 7\times 7\times 4$ mm$^3$, and therefore, field
inhomogeneities due to demagnetization are negligible.
The samples were field-cooled from above $T_c$ down to 1.6\,K in a field of
50\,mT and measured as a function of temperature (on the GPS spectrometer).
Additional measurements were performed down to $T \simeq 0.29$\,K (GPD
spectrometer; $^3$He cryostat) and $T \simeq 0.03$\,K (LTF spectrometer;
$^3$He/$^4$He dilution cryostat) in an applied field of 50\,mT. Measurements
in a series of fields ranging from 10\,mT to 640\,mT at 1.7\,K were also
performed. Typical counting statistics were $\approx 6 \times 10^6$ positron
events per each data point.

Isothermal magnetization loops at 1.85\,K were also recorded on the SQUID
magnetometer. In order to reduce demagnetization effects for these measurements
splinters of the samples were glued to a
quartz capillary with their longest dimensions parallel to the field
direction. Specific heat capacity as well as electrical resistance
measurements (ac, 93\,Hz, current density $j$ = 0.0072\,A\,mm$^{-2}$) at
$T_c$ and up to 320\,K were performed in magnetic fields up to 2.0\,T in a
measurement system (PPMS 9, Quantum Design).


\section{Analysis, results, and discussion}

\subsection{Muon spin rotation data}

Figure \ref{fig:Asy} exhibits typical $\mu$SR time spectra measured above and
below $T_c = 6.95$\,K in BaPt$_3$AuGe$_{12}$. The spectra of the other samples
are similar. Negligibly small muon relaxations above the respective $T_c$ are
observed in all samples for the whole field range. A fit with the function
$A\cos(\gamma_{\mu}Bt + \phi)\exp(-1/2 \sigma^2_N t^2)$ results in
$\sigma_N \le 0.06\,\mu$s$^{-1}$ (here, $A$, $B$, $\phi$, and $\sigma_N$ are
asymmetry, internal field, muon-spin phase, and relaxation rate,
respectively). The relaxation rate $\sigma_N < 0.06\,\mu$s$^{-1}$ is mostly
due to the nuclear magnetism of Ba, Pt, Au, and Ge isotopes, which causes a
weak depolarization of the muon-spin ensemble. Below $T_c$, all samples
exhibit relaxing $\mu$SR asymmetry spectra due to the spatial variation of the
internal field in the vortex-lattice state induced by the SC condensate. The
Fourier transform (FT) of this signal directly shows the field distribution
probed by the muon spins.

\begin{figure}[tb]
\includegraphics[width=\linewidth]{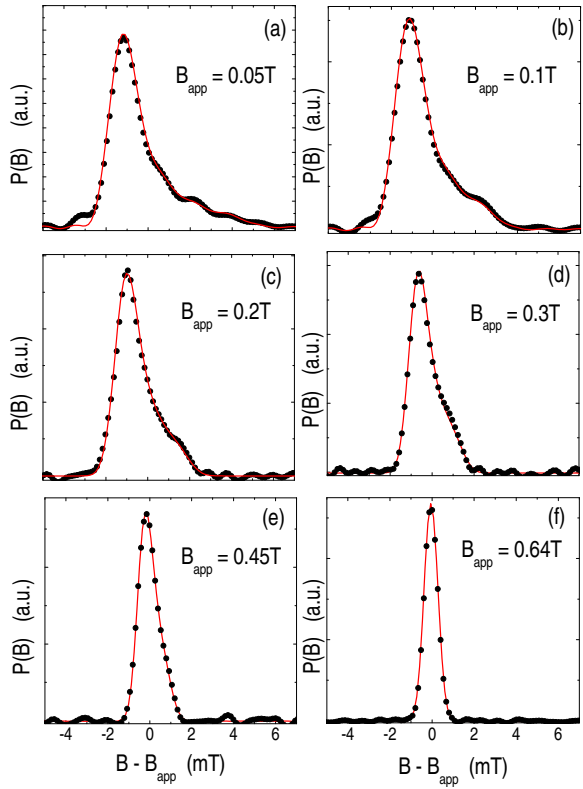}
\caption{(Color online) a--f: Fourier transforms (FT) of the $\mu$SR
time-spectra at 1.7\,K in different applied magnetic fields $\langle B\rangle
= 0.05$, 0.1, 0.2, 0.3, 0.45, and 0.64\,T for BaPt$_{3.5}$Au$_{0.5}$Ge$_{12}$
(corresponding reduced fields are $b = \langle B\rangle/B_{c2} = 0.06$, 0.12,
0.24, 0.36, 0.54, and 0.76). In this compound, fields close to
$B_{c2}$(1.7\,K) = 0.84(4)\,T are reachable. The solid lines are the FT of the
fit curves with Eq.\ (\ref{eq:Pt}). The field-dependent spectra are well
described by the field-independent parameters $\lambda = 239(4)$\,nm and $\xi
= 18.8(5)$\,nm.}
\label{fig:FT}
\end{figure}

Figure \ref{fig:FT} exhibits the FT spectra in BaPt$_{3.5}$Au$_{0.5}$Ge$_{12}$
in a broad range of fields. The asymmetric character of the vortex-lattice
field distributions -- reflecting the signatures of singularities at the
minimum, saddle, and core fields -- is clearly visible. Consequently, we
analyzed the $\mu$SR spectra for all BaPt$_{1-x}$Au$_x$Ge$_{12}$ samples using
the exact solution of the GL equations with the method suggested by Brandt
(see Appendix).\cite{Brandt03,Brandt97_NGLmethod} The spatial magnetic field
distribution $B(\mathbf{r}) = B(\mathbf{r}, \lambda, \xi, \langle B \rangle)$
within the unit cell of the flux-line lattice (FLL) was obtained by
minimization of Eq.\ (\ref{eq:GLfunctional}). From the obtained
$B(\mathbf{r})$ the probability field distribution for the ideal (defect-free)
FLL $P_\text{id}(B)$ is calculated as follows:

\begin{figure}[tb]
\includegraphics[width=\linewidth]{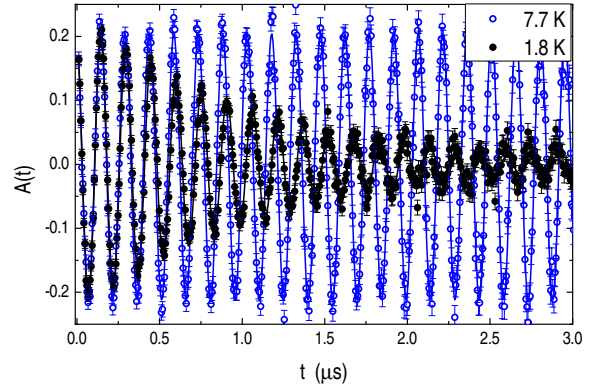}
\caption{(Color online) $\mu$SR time-spectra below and above $T_c = 6.95$\,K
in BaPt$_3$AuGe$_{12}$. The strong relaxation of the signal at 1.8\,K is due
to the formation of the flux-line lattice. Solid lines are fits to Eq.\
(\ref{eq:Pt}).}
\label{fig:Asy}
\end{figure}

\begin{equation}\label{eq:PBdefinition}
P_\text{id}(B') = \int \delta(B'-B(\mathbf{r}))d\mathbf{r}
\end{equation}

By assuming the internal field distribution $P_\text{id}(B)$ given by Eq.\
(\ref{eq:PBdefinition}) and accounting for the FLL disorder by multiplying
$P_\text{id}(B)$ to a Gaussian function,\cite{BrandtJLTPhys88and77} one
obtains the theoretical polarization function $P(t)$ given by:
\begin{align}\label{eq:Pt}
P(t) = &
A \exp\left[-\tfrac{1}{2}(\sigma_g^2+\sigma_N^2)t^2\right]
\int P_\text{id}(B)\cos({ \gamma_\mu B t} + \phi)dB \nonumber \\
+ & A_\text{bg} \exp\left(-\tfrac{1}{2}\sigma_\text{bg}^2t^2\right)
\cos(\gamma_\mu B_\text{bg}t+\phi),
\end{align}
which was used to fit the $\mu$SR time-spectra. Here, $\gamma_\mu = 2\pi \cdot
135.53$\,MHz/T is the muon gyromagnetic ratio, $A$ is the asymmetry of the
sample signal, $\phi$ is the phase of the muon-spin ensemble, $\sigma_g$ is a
parameter related to FLL disorder,\cite{Riesman95} and $\sigma_N$ the
additional muon depolarization due to the nuclear magnetism of various ions in
the samples. The parameters $A_\text{bg}$, $\sigma_\text{bg}$, and
$B_\text{bg}$ correspond to asymmetry, relaxation, and field of the background
signal, respectively. The asymmetries $A$ and $A_\text{bg}$ are found to be
temperature independent and $A+A_\text{bg} = 0.20$ (for the GPS and LTF
spectrometers) and $A+A_\text{bg} = 0.27$ (for the GPD spectrometer). The
background asymmetry $A_\text{bg} \simeq 0.004$ is negligibly small for the
measurements in the temperature range above 1.7\,K (on the GPS spectrometer)
while it is substantial in the measurements in the low-temperature range
($A_\text{bg} \approx 0.07$ on the LTF spectrometer and $A_\text{bg} \approx
0.20$ on the GPD spectrometer). The magnitude of $\sigma_N \simeq
0.05(1)$\,$\mu$s$^{-1}$ in BaPt$_{1-x}$Au$_x$Ge$_{12}$ was determined from
data above $T_c$. Zero-field $\mu$SR measurements in BaPt$_4$Ge$_{12}$ and
LaPt$_4$Ge$_{12}$\cite{Maisuradze10PRB} show that the ZF relaxation rate is
small and temperature independent, confirming the absence of magnetism. Thus,
$\sigma_N$ is negligibly small in comparison to the muon depolarization caused
by the nanoscale field inhomogeneities of the FLL. The background relaxation
is also small ($\sigma_\text{bg} \simeq 0.30$ or 0.007\,$\mu$s$^{-1}$), since
it corresponds to the signals originating from the copper or silver
sample-holder and from the walls of the cryostat.

The whole temperature dependence was fitted globally with Eq.\ (\ref{eq:Pt})
with the common parameters $A$, $A_\text{bg}$, $B_\text{bg}$,
$\sigma_\text{bg}$, and $\sigma_N$. In addition, the GL parameter $\kappa =
\lambda/\xi$ was taken as temperature-independent [{\textit i.e}.\ the
temperature dependent parameters $\xi = \lambda/\kappa$ and $B_{c2} =
\Phi_0/2\pi\xi^2$ are related to $\lambda(T)$]. The only
temperature-dependent parameters are $\lambda$ and $\langle B\rangle$. The
parameter $\sigma_g$ can be left free, however, relating $\sigma_g =
a/\lambda^2$ with the single global parameter $a$ reduces the total number of
parameters, thus reducing the error-bars for $\lambda$. Such a relation
corresponds to a rigid FLL.\cite{Riesman95} For a more detailed description
of the fitting procedure we refer to Ref.\ \onlinecite{Maisuradze08}.

\begin{figure}[tb]
\includegraphics[width=0.90\linewidth]{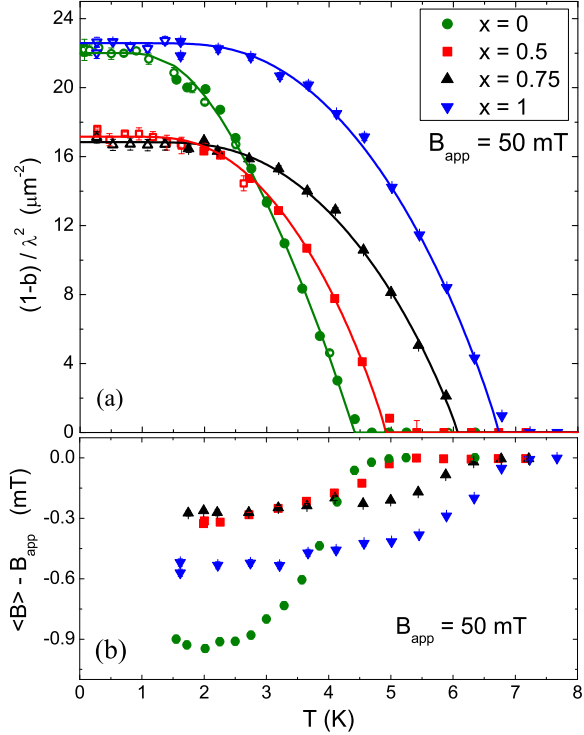}
\caption{(Color online) a: Temperature dependence of $(1-b)/\lambda^2 \propto
\rho_s$ measured at $B_\text{app} = 0.05$\,T in BaPt$_{4-x}$Au$_x$Ge$_{12}$
for $x = 0$, 0.5, 0.75, 1. All compounds exhibit exponential saturation of
$\rho_s$ in the low temperature limit, documenting a fully developed gap on
the Fermi surface. Solid symbols correspond to measurements above 1.6\,K (on
the GPS spectrometer) while the empty symbols correspond to those measured in
the low temperature limit (on the GPD at $x = 0.5$, 0.75, and 1 and on the
LTF at $x = 0$ spectrometers). The solid lines are fits to the Eq.\
(\ref{eq:SwaveSfld}). b: Temperature dependence of $\langle B\rangle$ for the
samples measured on the GPS spectrometer. Field inhomogeneity due to
demagnetization effects are only a small fraction of $\langle B \rangle -
B_\text{app}$.} \label{fig:SflD}
\end{figure}

The mean value of the superfluid density is related to the magnetic
penetration depth as follows (see Appendix): $\rho_s \propto (1-b)/\lambda^2
= 1/\tilde{\lambda}^2$.\cite{Brandt97_NGLmethod, Maisuradze09PRL} Here, $b =
\langle B\rangle/B_{c2}(0)$ is the reduced field and $\tilde{\lambda}$ is the
effective magnetic penetration depth.\cite{BrandtJLTPhys88and77} The
temperature dependencies of $1/\tilde{\lambda}^2$ for the
BaPt$_{4-x}$Au$_{x}$Ge$_{12}$ samples are shown in Fig.\ \ref{fig:SflD}a. The
superfluid density saturates exponentially below $\approx T_c/3$. This
documents the absence of quasiparticle excitations in the low-temperature
limit, which in turn suggests a superconducting gap without nodes in these
compounds. In Fig. \ref{fig:SflD}b we show fitting results for $\langle
B\rangle$. The magnitude of field inhomogeneity due to demagnetization
effects is only a small fraction of $\langle B\rangle-B_{\text app}$ since
shape of each sample is close to an ellipsoid (where internal field
is homogeneous).

\begin{figure}[tb]
\includegraphics[width=0.90\linewidth]{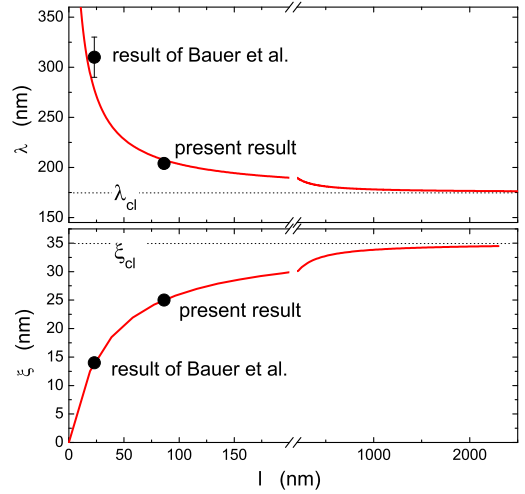}
\caption{(Color online) Magnetic penetration depth $\lambda$ and coherence
length $\xi$ of BaPt$_4$Ge$_{12}$ obtained in present study and by Bauer
\textit{et al.}\cite{Bauer07} (circles). The solid lines are fits to the data
using Eqs. (\ref{eq:DirtyXi})-(\ref{eq:DirtyLambda}) as described in the
text.}
\label{fig:XiLvsPath}
\end{figure}

The low-temperature limit of magnetic penetration depth and upper critical
field obtained for BaPt$_4$Ge$_{12}$ are $\lambda = 204(4)$ nm and $B_{c2} =
0.46(3)$ T (at $T$ = 1.7\,K), respectively. These values substantially differ
from these reported in Ref.\ \onlinecite{Bauer07} $\lambda = 320$ nm and
$B_{c2}$ = 1.8\,T, obtained by magnetization and specific heat measurements,
respectively. This discrepancy is explained by substantial scattering of
Cooper pairs on nonmagnetic impurities with the mean free path $l$ comparable
to the clean-limit coherence length $\xi_\text{cl}$. Indeed, the coherence
length and the magnetic penetration depth are related to these of the clean
limit ($l\rightarrow \infty$) as follows:\cite{Tinkham}

\begin{equation}\label{eq:DirtyXi}
\xi = \frac{\xi_\text{cl}l}{\xi_\text{cl}+l},
\end{equation}
\begin{equation}\label{eq:DirtyLambda}
\lambda = \lambda_\text{cl}\sqrt{1+\frac{\xi_\text{cl}}{l}}.
\end{equation}

Fitting this equation to the values of $\xi_1$ and $\xi_2$ reported in Ref.\
\onlinecite{Bauer07} and obtained here, respectively, with the additional
condition $l_1/l_2 = \rho^0_2/\rho^0_1 = 3.75$ (here $\rho^0_i$ are
corresponding residual resistivities) we obtain for $l_1 = 23$\,nm, $l_2 =
86$\,nm and $\xi_\text{cl} = 35$\,nm (see Fig.\ \ref{fig:XiLvsPath}). Note,
here we used the GL relation $B_{c2} = \Phi_0/2\pi \xi^2$ to obtain $\xi_i$
($i = 1$, 2). These values of mean free paths ($l_1$ and $l_2$) explain well
also different reports for magnetic penetration depths $\lambda_1 = 320$\,nm
and $\lambda_2 = 204$\,nm (see Fig.\ \ref{fig:XiLvsPath}). Consequently, the
compound BaPt$_4$Ge$_{12}$ is in between of clean and dirty-limit
superconductors. The residual resistivities of the compounds with $x = 0.5$,
0.75, and 1 are larger than for $x = 0$. Therefore, they are also dirtier
than the BaPt$_4$Ge$_{12}$ compound without Au substitution.

For the analysis of the superfluid density we adopt the BCS $s$-wave model
with arbitrary impurity scattering rate $1/\tau$:\cite{Xu95}
\begin{equation}\label{eq:SwaveSfld}
\frac{1}{\lambda^2} =
\frac{1}{\lambda_0^2}\pi k_BT\sum_{n=-\infty}^{\infty}\frac{1}{Z_n}
\frac{\Delta^2(T)}{[\epsilon_n^2+\Delta^2(T)]^{3/2}},
\end{equation}
with
\begin{equation}
Z_n = 1+\frac{\hbar}{\tau}\frac{1}{\sqrt{\epsilon_n^2+\Delta^2(T)}}.
\end{equation}
Here, the classical BCS temperature dependence of the gap was used $\Delta(t)
= \Delta_0\delta(t)$ with $\delta(t) = \tanh\{1.82[1.018(t - 1)^{0.51}]\}$
(with $t = T/T_c$).\cite{Tinkham} $k_B$ and $\hbar$ are Boltzmann and reduced
Planck constants, respectively, $\epsilon_n = \pi T(2n+1)$ are Matsubara
frequencies while $Z_n$ are renormalization factors for $\epsilon_n$ and the
superconducting gap $\Delta$. In the extreme cases of the clean
($\tau\rightarrow \infty$) and dirty ($\tau\rightarrow 0$) limits this
equation converges to the classical clean and dirty superconductor curves
(see Appendix).\cite{Tinkham} For the Fermi velocity\cite{Bauer07} $v_F =
52000$\,m\,s$^{-1}$ and mean free path $l = 86$\,nm of BaPt$_4$Ge$_{12}$ we
obtain the scattering time $\tau = 1.6\times10^{-12}$\,s.

The fits of Eq.\ (\ref{eq:SwaveSfld}) to $1/\tilde{\lambda}^2$ are shown in
Fig.\ \ref{fig:SflD}a by solid lines. The fit results for $\Delta_0$, $T_c$,
and the low temperature limits of $1/\tilde{\lambda}(0)^2$ are summarized in
Table \ref{tab:aSFLt}. As can be seen in the Appendix, there is a correlation
between the parameters $\tau$ and $\Delta_0$ in Eq.\ (\ref{eq:SwaveSfld}).
Therefore, for BaPt$_4$Ge$_{12}$ ($x = 0$) we used $\tau =
1.6\times10^{-12}$\,s as estimate. For the compounds with $x = 0.5$, 0.75,
and 1 we used the upper limit of $\tau_\text{max} = 1.6\times10^{-12}$\,s,
since they are dirtier than the BaPt$_4$Ge$_{12}$ compound. The fit of the
data was performed for $\tau = \tau_\text{max}$ and $\tau\rightarrow 0$
(dirty limit, when $\tau \ll \xi_\text{cl}/v_F$). Thus, we obtain the upper
and lower limits of $\Delta_0^\text{max}$ and $\Delta_0^\text{min}$,
respectively. The values of the gap reported in Table \ref{tab:aSFLt} are
$\Delta_0 = 0.5(\Delta_0^\text{max}+\Delta_0^\text{min})$ with errors
including the uncertainty in $\tau$ and the (much smaller) statistical error.

With increasing Au substitution the $T_c$ increases, however the gap-to-$T_c$
ratio $\Delta_0/k_BT_c$ increases more suddenly and remains essentially
unchanged for the Au substitutions $x = 0.5$, 0.75, and 1. The $T_c$ of
phonon-mediated SC may be described within the McMillan formula.\cite{Tinkham}
The Debye temperature is practically the same for our four
compounds.\cite{Gumeniuk08AuOpt} Thus, the only factor determining the
increase of $T_c$ is the electronic DOS at the Fermi level, which
significantly increases with Au substitution beyond $x = 0.4$ (cf.\ Fig.\ 1 in
Ref.\ \onlinecite{Gumeniuk08AuOpt}).

Another interesting feature is the dependence of the superfluid density upon
Au substitution. With increasing $x$, $\rho_s(0)$ first decreases, goes
through a minimum, and then increases with further increasing $x$. Such a
behavior is rather unusual and contrasts with previous observations of a
power-law-like relation between $T_c$ and the superfluid density in cuprate
high-$T_c$ superconductors,\cite{Uemura89PRL} NbB$_2$,\cite{Khasanov08NbB2}
MgB$_2$,\cite{Serventi05MgB2} or predicted
theoretically.\cite{Kim91,Schneider07} However, in the present case such a
behavior can be understood. The superfluid density and the magnetic
penetration depth $\lambda$ are dependent on the scattering rate of the Cooper
pairs $\tau$ (or the mean free path $l$) [see Eq.\ \ref{eq:DirtyXi}].
Therefore, the minimum in $\rho_s(0)$ is probably due to the dependence of
$\tau$ on the Au content $x$.

\begin{table}[tb]
\caption{\label{tab:aSFLt} Summary of fit results with Eq.\
(\ref{eq:SwaveSfld}) for $T_c$ and $\Delta_0$ in BaPt$_{4-x}$Au$_x$Ge$_{12}$,
$x = 0$, 0.5, 0.75, 1. In addition, the low-temperature limit of
$1/\tilde{\lambda}^2$ and the gap-to-$T_c$ ratio are given.}
\begin{center}
\begin{tabular}{ccccccc}
\hline \hline
$x$    & $\Delta_0$ & $T_c$   & $\tilde{\lambda}^{-2}(0)$ & $\Delta_0/k_BT_c$ \\
       & (meV)      & (K)     & ($\mu$m$^{-2}$)           &                   \\
\hline
0      & 0.65(1)    & 4.45(3) & 22.0(1)                   &  1.70(3)          \\
0.5    & 0.88(4)    & 4.93(1) & 17.15(7)                  &  2.07(8)          \\
0.75   & 1.13(5)    & 6.10(4) & 16.84(9)                  &  2.15(6)          \\
1      & 1.20(5)    & 6.86(5) & 22.6(3)                   &  2.02(9)          \\
\hline\hline
\end{tabular}
\end{center}
\end{table}

\begin{table}[tb]
\caption{\label{tab:Lambda} Magnetic penetration depth ($\lambda$) and
coherence length ($\xi$) obtained from the global fit of the data in the
field range of 0.05 to 0.64\,T at 1.7\,K. In addition, the values for
$B_{c2}$ at $T$ = 1.7\,K, Ginzburg-Landau parameter $\kappa = \lambda/\xi$,
and residual resistivity $\rho_0$ are listed.}
\begin{center}
\begin{tabular}{ccccccccccc}
\hline \hline
$x$    & $\lambda$ & $\xi$    & $\kappa$ & $B_{c2}$(1.7\,K) & $\rho_0$          \\
       & (nm)      & (nm)     &          & (T)              & ($\mu\Omega$\,cm) \\
\hline
0      & 204(4)    & 25.5(15) & 8.0(5)   &  0.46(3)         & 15.1              \\
0.5    & 239(4)    & 18.8(9)  & 12.7(7)  &  0.84(4)         & 33.6              \\
0.75   & 240(4)    & 15.0(9)  & 16.0(9)  &  1.68(5)         & 36.8              \\
1      & 210(4)    & 13.4(8)  & 15.7(9)  &  1.93(5)         & 31.5              \\
\hline \hline
\end{tabular}
\end{center}
\end{table}

Further information about the order parameter can be obtained from the field
dependence of the superfluid density. As it is known for superconductors with
nodes in the gap, a significant field dependence of $\lambda$ is
observed,\cite{Amin00} while for the large number of classical BCS
superconductors $\lambda$ is field-independent.\cite{LandauKeller07} A fit of
the $\mu$SR-time spectra at different fields and for $T = 1.7$\,K using Eq.\
(\ref{eq:Pt}) results in field-independent values of
$\lambda$.\cite{Footnote1} For the fit the field-independent value of $\xi$
obtained from the GL relation $B_{c2} = \Phi_0/2\pi\xi^2$ was used.
Therefore, we next fitted the whole field dependence of the spectra globally
with common parameters $\lambda$ and $\xi$. Note that the values of the GL
parameter $\kappa = \lambda/\xi$ used in the fit of the temperature
dependence were obtained from the fit of the corresponding field scan. In
Fig.\ \ref{fig:FT} we show the FT $\mu$SR spectra for the BaPt$_3$AuGe$_{12}$
compound at $T$ = 1.7\,K for the broad range of reduced fields $b = B/B_{c2}
= 0.06$, 0.12, 0.24, 0.36, 0.54, and 0.76. The fit results in
field-independent values of the magnetic penetration depth ($\lambda$ =
239(4)\,nm) and of the coherence length ($\xi = 18.8(5)$\,nm). This value of
$\xi$ is in good agreement with that obtained by the GL relation from the
corresponding $B_{c2}$ ($B_{c2} = \Phi_0/2\pi\xi^2$). The fit results for
$\lambda$, $\xi$, and $\kappa$ in BaPt$_{4-x}$Au$_x$Ge$_{12}$ for $x = 0$,
0.5, 0.75, and 1 are summarized in Table \ref{tab:Lambda}. The upper critical
field for $x = 1$ is in fair agreement with the value given in our previous
work,\cite{Gumeniuk08AuOpt} but for BaPt$_4$Ge$_{12}$ the $B_{c2}(0)$ is much
lower than previously reported.\cite{Bauer07,Gumeniuk08AuOpt} This drastic
discrepancy is investigated and discussed in the following section.

\subsection{Macroscopic measurement data}

Previously, for BaPt$_4$Ge$_{12}$ an upper critical field $B_{c2}(0)$ of about
2.0\,T was reported by our group, mainly based on $T_c(B_\text{app})$ data
from resistivity measurements in fixed applied fields
$B_\text{app}$.\cite{Gumeniuk08AuOpt} This value was confirmed by similar data
of Bauer \textit{et al.}\cite{Bauer07} [$B_{c2}(0)$ = 1.8\,T] from both
resistivity \textit{and} specific heat data in field.\cite{Bauer07} However,
the $T_c$ reported for BaPt$_4$Ge$_{12}$ in Ref.\ \onlinecite{Bauer07} is
5.35\,K (from both resistivity \textit{and} specific heat), which is
inconsistent with the magnetic onset $T_c \approx 4.9$\,K of our present and
$T_c \approx 5.0$\,K of our previous $x = 0$ samples. The origin of the
drastically different upper critical field values as well as of the unusually
large variation of $T_c$ for BaPt$_4$Ge$_{12}$ samples remained unclear. For
this reason we (re-)investigated the present BaPt$_4$Ge$_{12}$ ($x = 0$)
sample as well as the $x = 0$ and $x = 1$ samples from our previous study
\cite{Gumeniuk08AuOpt} by macroscopic probes (magnetization, specific heat,
electrical resistivity).

\begin{figure}[tb]
\includegraphics[angle=90,width=0.95\linewidth]{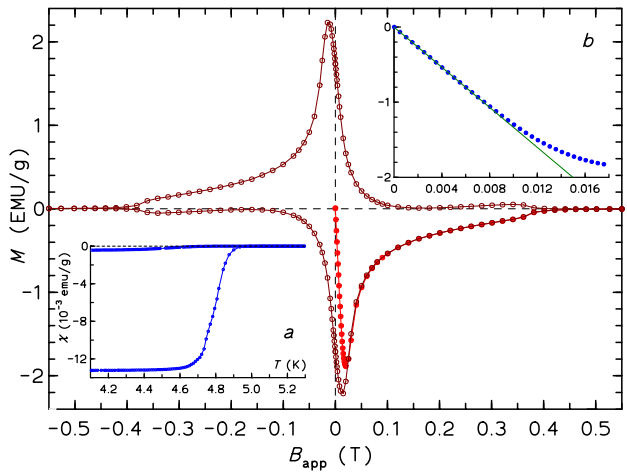}
\caption{(Color online) Magnetization loop of BaPt$_4$Ge$_{12}$ at $T$ =
1.85\,K up to $B_\text{max} = \pm$2.0\,T. The initial curve is marked by
(light-red) full circles, the other segments by (dark-red) open circles. Inset
a: Zero-field cooled (Meissner effect) and field-cooled (shielding)
susceptibility in a nominal field of 2\,mT. Inset b: low-field magnetization
showing the deviation from the initial linear behavior (straight line) at
$B_{c1}$.}
\label{fig:Mag1}
\end{figure}

\begin{figure}[tb]
\includegraphics[angle=90,width=0.95\linewidth]{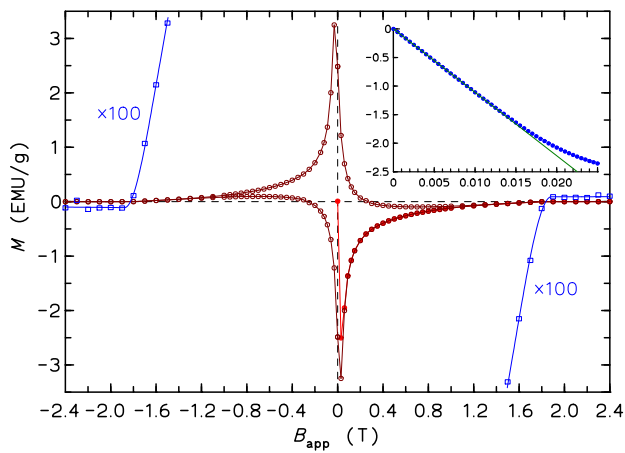}
\caption{(Color online) Magnetization loop of BaPt$_3$AuGe$_{12}$ at $T$ =
1.85\,K up to $B_\text{max} = \pm$3.0\,T. The (blue) squares (with a line as
guide to the eye) show a 100-fold magnification of the data close to
$B_{c2}$. Inset: low-field magnetization showing the deviation from the
initial linear behavior (straight line) used for estimating $B_{c1}$.}
\label{fig:Mag2}
\end{figure}

An isothermal magnetization loop at $T = 1.85$\,K for the BaPt$_4$Ge$_{12}$
sample used for the $\mu$SR measurements is given in Fig.\ \ref{fig:Mag1}. It
shows the typical picture of a type II superconductivity with medium large GL
parameter $\kappa$. A weak second peak effect in $M(B_\text{app})$ is
observed around 0.35\,T, indicating relatively weak flux-line pinning in a
pure sample.\cite{Nb-crystals} Above this field the hysteresis drastically
diminishes (this field is often taken as upper critical field
$B_{c2}$)\cite{Nb-crystals} and becomes reversible above $B_\text{app}$ =
500(30)\,mT. The reversible SC magnetization decreases to a value of less
than 1/1000 of the maximum magnetization signal (the noise level of the
measurement) at $B_\text{app}$ = 540(50)\,mT, which we adopt as the upper
critical field $B_{c2}$(1.85\,K). Considering the only slightly different
temperatures, this value is compatible with $B_{c2}(0)$ from our $\mu$SR
investigation (see Table \ref{tab:Lambda}). The lower critical field can be
estimated from the first deviation of $M(B_\text{app})$ from linearity (Fig.\
\ref{fig:Mag1}b). Adopting a 0.5\,{\%} criterion for a significant deviation,
we find $B_{c1} = 6.0(1.0)$\,mT, again for $T = 1.85$\,K. This experimental
value for $B_{c1}$ is, however, only a lower limit due to the strong
influence of the demagnetization effect for a nearly perfect diamagnet. From
the $B_{c2}$ value determined from magnetization and $\kappa = 8.0$ from
Table \ref{tab:Lambda} we obtain with the relation $B_{c2} = \sqrt{2}\kappa
B_\text{c,th}$ a thermodynamic critical field $B_\text{c,th} \approx 48$\,mT,
which is clearly larger than the value calculated from the free enthalpy
difference from specific heat ($\approx 40$\,mT for $T$ = 1.85\,K). Using
$B_{c1} \approx (\ln{\kappa} + 0.5) B_\text{c,th}/2\kappa$ (valid for small
$\kappa$)\cite{Brandt03} we find $B_{c1} = 7.7$\,mT, in fair agreement with the estimate
from the magnetization curve. Only slightly different values for the critical
fields were obtained from similar magnetization data (not shown) taken on the
BaPt$_4$Ge$_{12}$ sample used in Ref.\ \onlinecite{Gumeniuk08AuOpt} ($B_{c1}
= 4.2(1.0)$\,mT, $B_{c2} = 590(50)$\,mT, both at $T = 1.85$\,K).

The magnetization curve for the BaPt$_3$AuGe$_{12}$ sample ($x = 1$) of the
present study is given in Fig.\ \ref{fig:Mag2}. There is no visible second
peak effect and, thus, the upper critical field can be determined accurately
from the sharp kinks in $M(B_\text{app})$ ($B_{c2} = 1820(20)$\,mT; see
100-fold magnification of the data in Fig.\ \ref{fig:Mag2}). The estimated
$B_{c1}$ is 11.5(1.0)\,mT ($T = 1.85$\,K; criterion 0.5\,{\%} deviation).
While the $B_{c2}$ value is only slightly lower compared to the one in Table
\ref{tab:Lambda} the GL parameter $\kappa \approx 13.1$ is clearly lower than
the value determined by $\mu$SR spectroscopy. The $B_\text{c,th}$ calculated
using $\kappa$ from Table \ref{tab:Lambda} and $B_{c2}$ from the
magnetization curve is $\approx 99$\,mT, which is again clearly larger than
the value derived from the specific heat data ($\approx 79$\,mT at 1.85\,K).

\begin{figure}[tb]
\includegraphics[angle=90,width=0.95\linewidth]{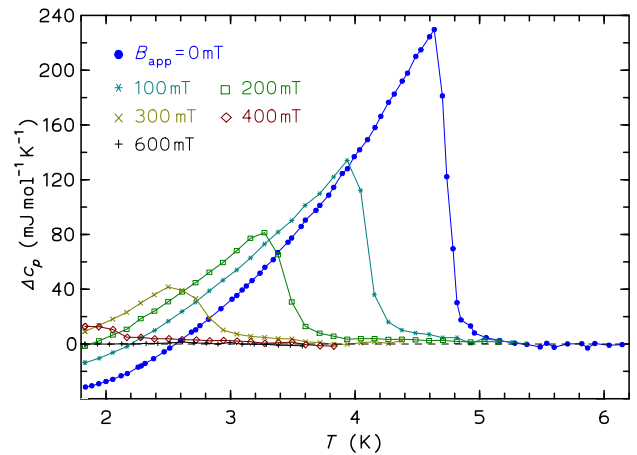}
\caption{(Color online) Difference of the specific heat $\Delta
c_p(T,B_\text{app})$ of BaPt$_4$Ge$_{12}$ (sample from $\mu$SR investigation)
between SC and normal state ($B_\text{app} = 2.0$\,T). $\Delta c_p = 0$ is
indicated by a dashed line.}
\label{fig:Cp}
\end{figure}

In Fig.\ \ref{fig:Cp} the difference specific heat $\Delta c_p(T) =
c_p(B_\text{app}) - c_p(B > B_{c2})$ is plotted for the present
BaPt$_4$Ge$_{12}$ sample. For fields $\geq 600$\,mT no SC signal is observed
above our lowest temperature of 1.8\,K. The midpoints of the
second-order-type transitions $T_c(B_\text{app})$ were evaluated. The
quadratic extrapolation of these data for $B_{c2}(T)$ to zero temperature
results in $B_{c2}(0) = 470(50)$\,mT, in excellent agreement with the values
from $\mu$SR and magnetization. For the sample used in Ref.\
\onlinecite{Gumeniuk08AuOpt} we extrapolate $B_{c2}(0) = 540(50)$\,mT.
Specific heat data in field for the other compositions are given in Ref.\
\onlinecite{Gumeniuk08AuOpt}.

\begin{figure}[tb]
\includegraphics[angle=90,width=\linewidth]{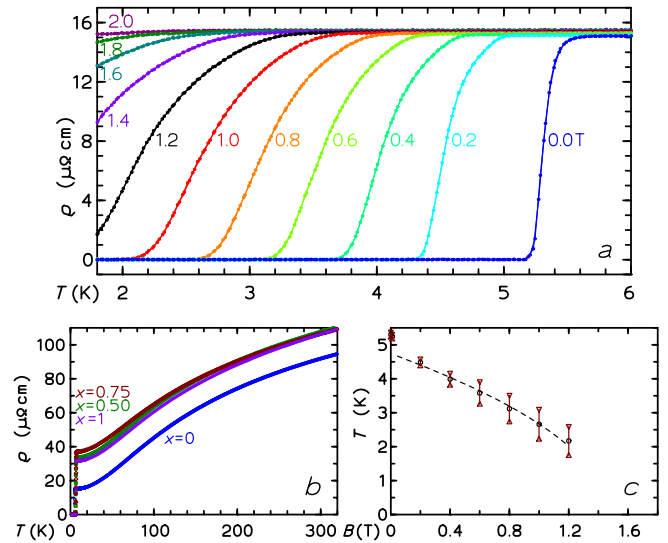}
\caption{(Color online) a: Electrical resistivity $\rho(T,B_\text{app})$
around $T_c$ of the present BaPt$_4$Ge$_{12}$ sample for different applied
fields. b: Electrical resistivity $\rho(T)$ in zero field for the $x = 0$,
0.50, 0.75, and 1 samples. c: Variation of the onset, mid (circles), and
zero-resistance temperatures with applied field. The dashed line is a
quadratic fit for $T_\text{c,mid}(B_\text{app})$.}
\label{fig:rho}
\end{figure}

The electrical resistivity of the present $x = 0$ sample at 300\,K is
$\approx 90\,\mu\Omega$m with a residual resistance ratio 6.1 (Fig.\
\ref{fig:rho}b). Such low RRR values are not typical for polycrystalline
samples of other $M$Pt$_4$Ge$_{12}$ compounds (cf.\ RRR = 33 or 42
[\onlinecite{Gumeniuk08}], RRR $\approx$ 100 [\onlinecite{Kaczorowski08}], or
RRR $\geq$ 100 [\onlinecite{Gumeniuk2011Ce}]). Obviously, the crystalline
quality of polycrystalline BaPt$_4$Ge$_{12}$ samples is worse compared to
that of other members of the family of filled Pt-Ge skutterudites. RRR
$\approx$ 6 however indicates that a BaPt$_4$Ge$_{12}$ sample with a clearly
lower defect concentration than in Refs.\ \onlinecite{Bauer07} or
\onlinecite{Khan08a} has been achieved. For the present sample the SC
transition in $\rho(T,B_\text{app})$ decreases continuous with increasing
field, except for very low fields (Fig.\ \ref{fig:rho}a). The onset, mid, and
zero-resistance temperatures are plotted against $B_\text{app}$ in Fig.\
\ref{fig:rho}c. Surprisingly, the transition in $\rho(T)$ is still complete
for a field of 1.0\,T and the onset is even visible at 1.9\,K in 1.8\,T. Such
high upper critical fields are in agreement with the conclusions in Refs.\
\onlinecite{Bauer07}, but in strong contrast to the consistently much lower
$B_{c2}$ values obtained from the bulk-probes $\mu$SR, magnetization, and
specific heat. A quadratic extrapolation of $T_\text{c,mid}$ (range of fit
0.2--1.2\,T) to zero temperature results in $B_{c2}^\text{res}(0) = 1460$\,mT
(dashed line in Fig.\ \ref{fig:rho}c). In addition, a clear anomaly is seen
for the lowest fields, where the resistive $T_c$ is almost 0.5\,K higher than
expected from the extrapolation curve. Actually, the extrapolated resistive
$T_\text{c,mid}(0) \approx 4.74$\,K agrees well with that from the bulk
measurements. For the sample of BaPt$_3$AuGe$_{12}$ no significant
discrepancy between bulk and resistive value for $B_{c2}$ are found (see
Ref.\ \onlinecite{Gumeniuk08AuOpt})

What is the origin of this discrepancy in the $B_{c2}$ and $T_c(0)$ values
from bulk properties and resistivity in BaPt$_4$Ge$_{12}$? In the low-field
susceptibility ($B_\text{app}$ = 2\,mT, $\mu$SR sample) a bulk $T_c$ onset of
4.87\,K is observed (determined by the tangent to the steepest slope of the
field-cooling Meissner transition). However, above $\approx 5.0$\,K there is
still a very weak diamagnetic signal, which vanishes exponentially with
increasing temperature. The signal is only little weaker in field-cooling than
measured after zero-field cooling (zfc), but the maximum of this zfc signal is
about 3 orders of magnitude lower than the zfc signal at 4.0\,K (or 1.5 orders
of magnitude weaker than the bulk Meissner effect).

This weak diamagnetism, the concomitant zero electrical resistance, and the
too large $B_{c2}$ value from resistivity data may root in two phenomena: i.
the presence of a minor SC impurity phase which forms a percolating SC network
with an about 0.5\,K higher $T_c(0)$ and much higher $B_{c2}$ than the main
phase, or, ii., strong classical surface superconductivity of the main phase
with a critical field $B_{c3}(0) \gg B_{c2}(0)$. The second possibility seems
to be unlikely due to the facts that $T_\text{c,bulk}(0) \neq
T_\text{c,surface}$ and that the required surface critical field $B_{c3}$
would well need to exceed the Saint-James--de Gennes limit of $\approx 1.7
B_{c2}$.\cite{SJdG1963}

For BaPt$_4$Ge$_{12}$ no homogeneity range is observed since the lattice
parameter of the filled-skutterudite phase in a sample with composition
Ba$_{0.9}$Pt$_4$Ge$_{12}$ is 8.6837(3)\,{\AA}, which is practically the same
as for BaPt$_4$Ge$_{12}$.\cite{Gumeniuk2010} Extended EDXS analyses on
metallographic polished surfaces of the present BaPt$_4$Ge$_{12}$ sample
result in a composition Ba$_{0.9(1)}$Pt$_{4.0(1)}$Ge$_{11.9(1)}$, which agrees
very well with the nominal one. Interestingly, there is a significant
difference of the lattice parameter of all our BaPt$_4$Ge$_{12}$ samples
(present sample $a$ = 8.6838(5)\,{\AA}) with that reported by Bauer \textit{et
al.}\cite{Bauer07} ($a$ = 8.6928(3)\,{\AA}),\cite{Bauer07} which we currently
cannot explain.

The currently studied large BaPt$_4$Ge$_{12}$ sample contains besides the
BaPt$_4$Ge$_{12}$ main phase also some BaPtGe$_3$ (no superconductivity
observed above 1.8\,K),\cite{Demchyna06a,Schnelleunpub}). The content of this
phase is estimated from Rietveld refinements to be about 4\,{\%}. Five weak
lines in the X-ray diffraction pattern belong to PtGe$_2$. These lines are too
weak to refine a phase content, therefore we estimate a PtGe$_2$ phase
fraction of below 2\,{\%}. PtGe$_2$ is reported to be a superconductor with
$T_c$ = 0.4\,K.\cite{MatthiasReview} The presence of these minority phases in
the BaPt$_4$Ge$_{12}$ sample thus also cannot explain the observation of a
higher upper critical field value in resistivity data.

The resistive percolation (a SC path) at a higher temperature than the bulk
$T_c$ hints at a modification of the surface layers of the grains of the
majority skutterudite phase, probably due to crystallographic defects or
strain. These effects will result in a larger scattering rate and a shorter
mean free path of the charge carriers, thus making the superconductor dirtier,
subsequently enlarging the effective penetration depth well above the bulk
value.

Since the present samples are polycrystalline pieces, an estimate of the mean
free path from the residual resistivity values by the standard
formula\cite{AshcroftMermin} is problematic and gives at best a lower limit
for $l$. Moreover, skutterudites are not simple metals which can be treated in
a one-band model. Our estimate of the minimal mean free path in
BaPt$_{4-x}$Au$_x$Ge$_{12}$ using a free-electron model results in
$l_\text{min} \approx25$, 14, 14, and 19\,nm for $x = 0$, 0.5, 0.75, and 1,
respectively.\cite{Remark-on-n} In view of these values of $l_\text{min}$, the
superconductivity in the bulk of the crystallites is neither in the clean nor
in the dirty limit. On the surface, however, the superconductivity seems to be
in the dirty limit, leading to much shorter coherence lengths than in the
bulk. Hence, crystalline defects or impurities on the grain surfaces probably
lead to the higher upper critical field value in resistivity data. An open
question is the clearly higher $T_c$ of these grain surfaces. The $T_c$ of a
superconductor with defects is -- in most cases -- lower than the $T_c$ of the
pure material, however, it is also known that strain, especially on surfaces,
can drastically enhance the $T_c$. While the growth of single crystals of
sufficient size of BaPt$_4$Ge$_{12}$ was not successful until now,
investigations on such crystals would be highly desirable.


\section{Conclusion}

We performed an investigation using transverse-field $\mu$SR spectroscopy for
a series of polycrystalline BaPt$_{4-x}$Au$_x$Ge$_{12}$ superconductors with
$x = 0$, 0.5, 0.75, and 1. Highly asymmetric $\mu$SR time spectra were
analyzed within the framework of the Ginzburg-Landau (GL) theory by precise
minimization of the GL free energy.\cite{Brandt97_NGLmethod} Zero-temperature
magnetic penetration depths [$\lambda(0)$] and GL parameters ($\kappa =
\lambda/\xi$) were evaluated (see Table \ref{tab:Lambda}). The temperature
dependence of the superfluid density $\rho_s$ in all the compounds saturates
exponentially in the low-temperature limit, which documents the absence of
nodes in the superconducting gap function. This finding is in agreement with
the results of a previous NMR study.\cite{Magishi09} Our analysis shows that
$\rho_s$ is well described within the classical $s$-wave BCS model
with gap-to-$T_c$ ratios ($\Delta_0/k_BT_c$) of 1.70, 2.07, 2.15, and 2.02 for
$x = 0$, 0.5, 0.75, and 1, respectively. These ratios are in fair agreement
with the reduced specific heat jump $\delta c_p/\gamma_NT_c$ from our previous
study.\cite{Gumeniuk08AuOpt} The observation of field-independent $\lambda$
values further supports the classical $s$-wave pairing scenario for these
compounds. Thus, the present experimental results from bulk probes point to
the classical $s$-wave phonon-mediated superconductivity for all compounds in
the series BaPt$_{4-x}$Au$_x$Ge$_{12}$ up to $x = 1$. The upper critical field
data from the $\mu$SR study are in good agreement with bulk-sensitive
thermodynamic measurements of the upper critical fields of BaPt$_4$Ge$_{12}$
and BaPt$_3$AuGe$_{12}$. The origin of much higher upper critical fields
observed in electrical resistivity measurements for the present
BaPt$_4$Ge$_{12}$ sample (as in previous
reports\cite{Bauer07,Gumeniuk08,Gumeniuk08AuOpt}) is due to a larger carrier
scattering rate at the surface of the crystallites.

\section*{Acknowledgements}

We would like to acknowledge fruitful discussions concerning GL theory with
the late E.\ H.\ Brandt. This work was partly performed at the Swiss Muon
Source (S$\mu$S), Paul Scherrer Institut (PSI, Switzerland). The work was
supported by the NCCR program \textit{Materials with Novel Electronic
Properties} (MaNEP) sponsored by the Swiss National Science Foundation. We
acknowledge helpful discussions with H.\ Rosner and thank Yu.\ Grin for
continuous support of our work.


\appendix*

\section{Details of calculations and GL definitions}

Below we describe some details of our calculations and introduce the basic
definitions of the Ginzburg-Landau (GL) theory used in this analysis. As
shown by Abrikosov, a type II superconductor forms a periodic vortex or
flux-line lattice (FLL) in a range of magnetic fields ($B$).\cite{Abrikosov}
Here, $B_{c1} < B < B_{c2}$, where $B_{c1}$ and $B_{c2}$ are the lower and
upper critical fields, respectively. The GL theory used by Abrikosov occurred
to be one of the most useful approaches for the evaluation of the field
distribution in a type II superconductor (although it is strictly valid only
close to $T_c$) and forms the basis for the analysis of transverse field (TF)
$\mu$SR data.

\begin{figure}[tb]
\includegraphics[width=0.49\linewidth]{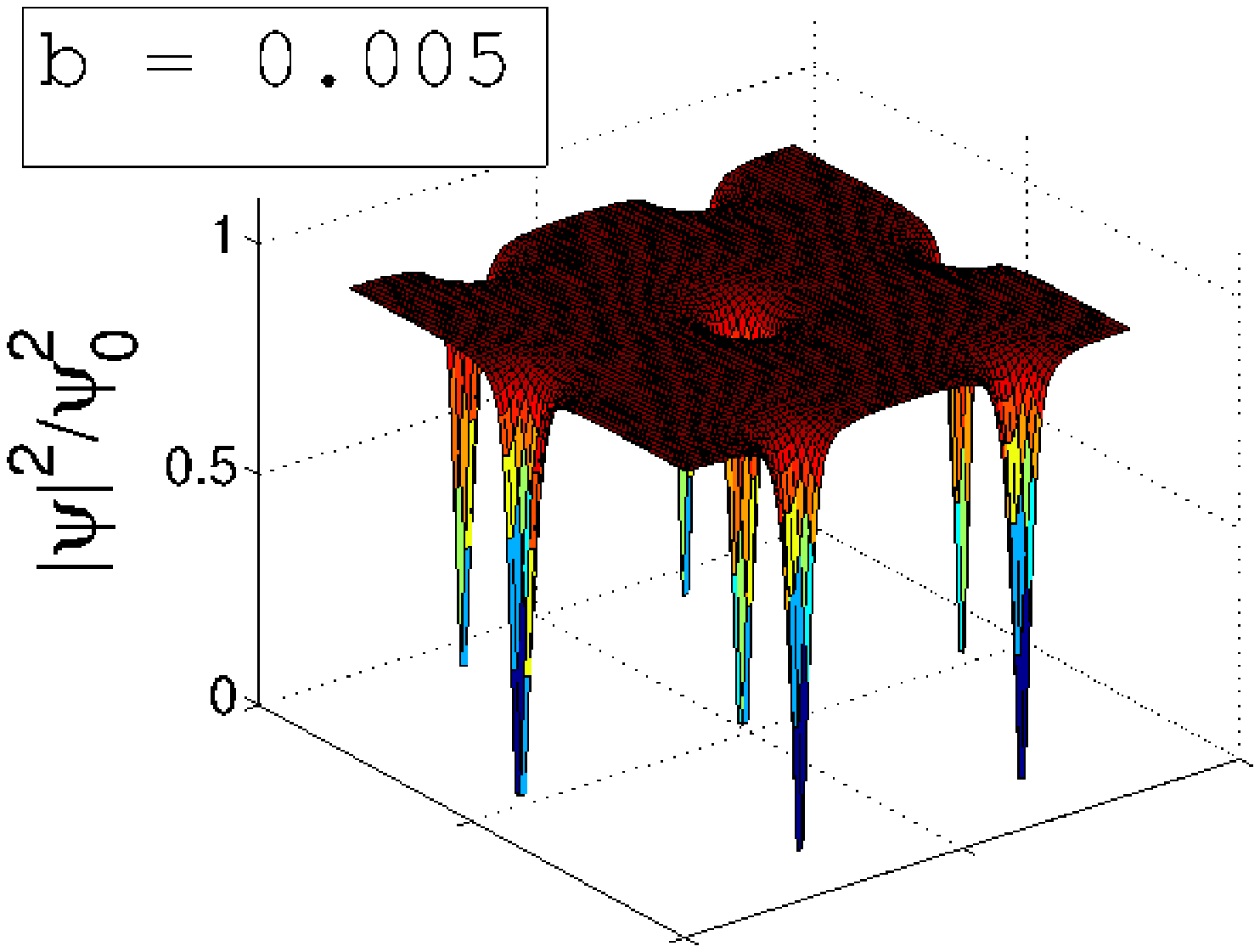}
\includegraphics[width=0.49\linewidth]{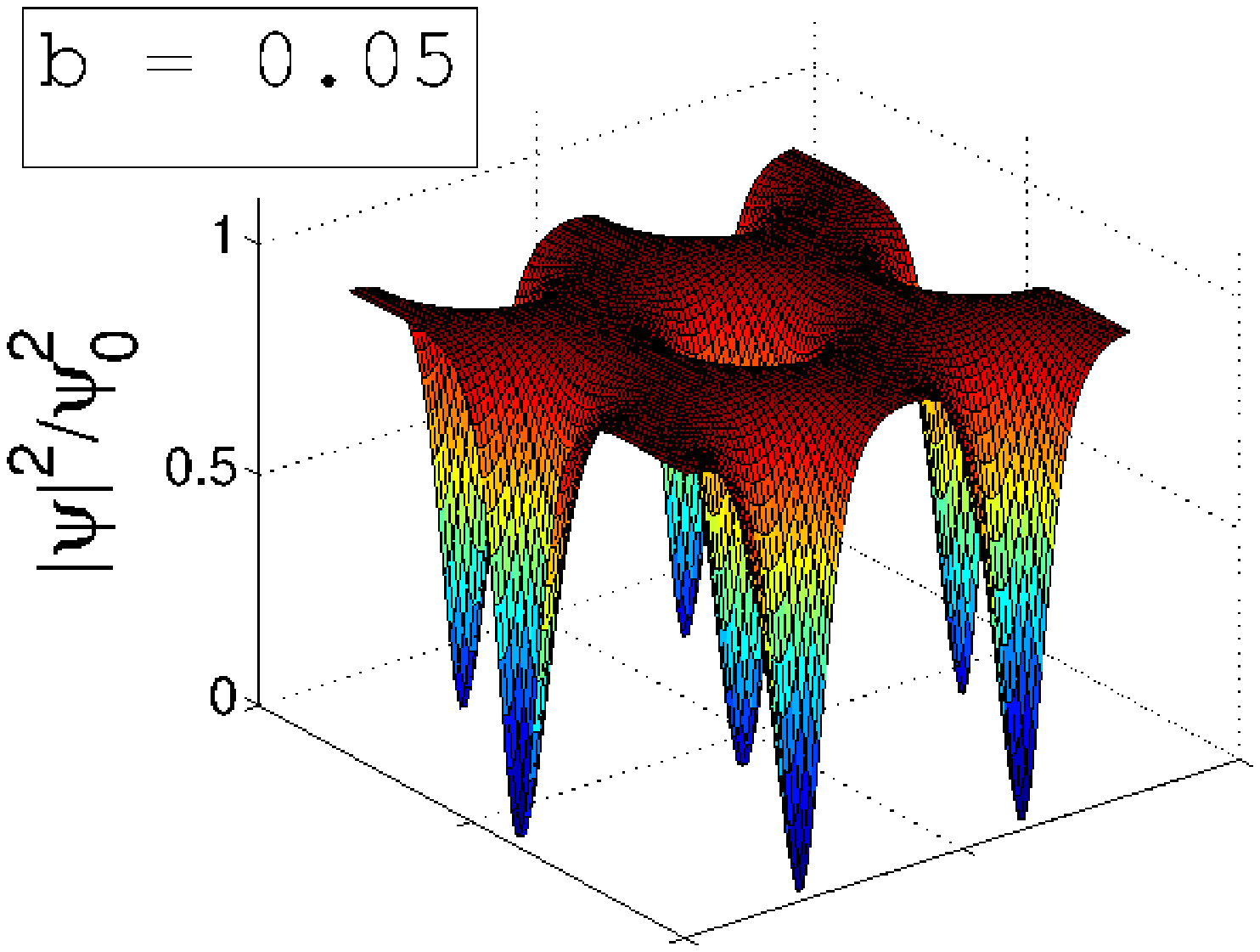}
\includegraphics[width=0.49\linewidth]{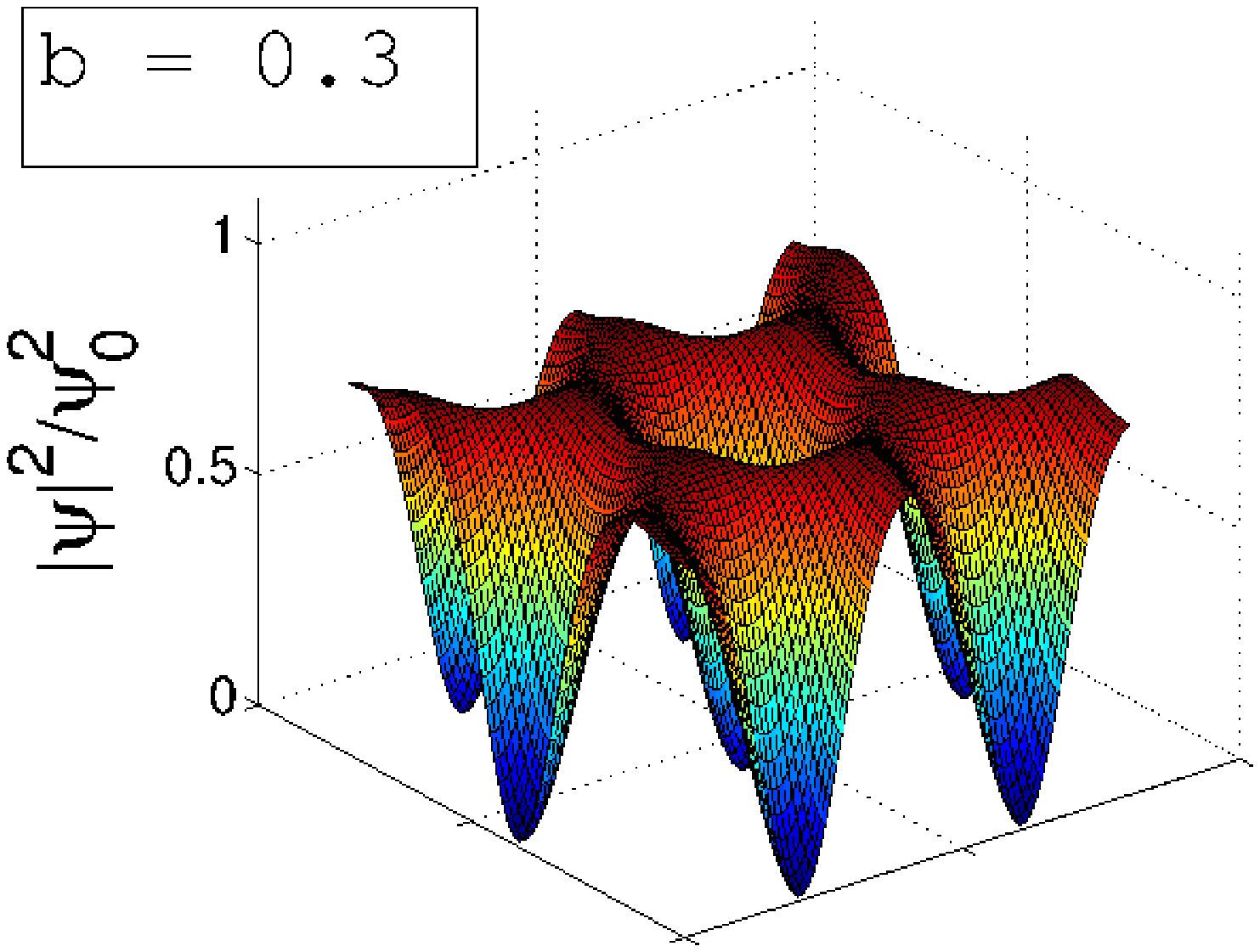}
\includegraphics[width=0.49\linewidth]{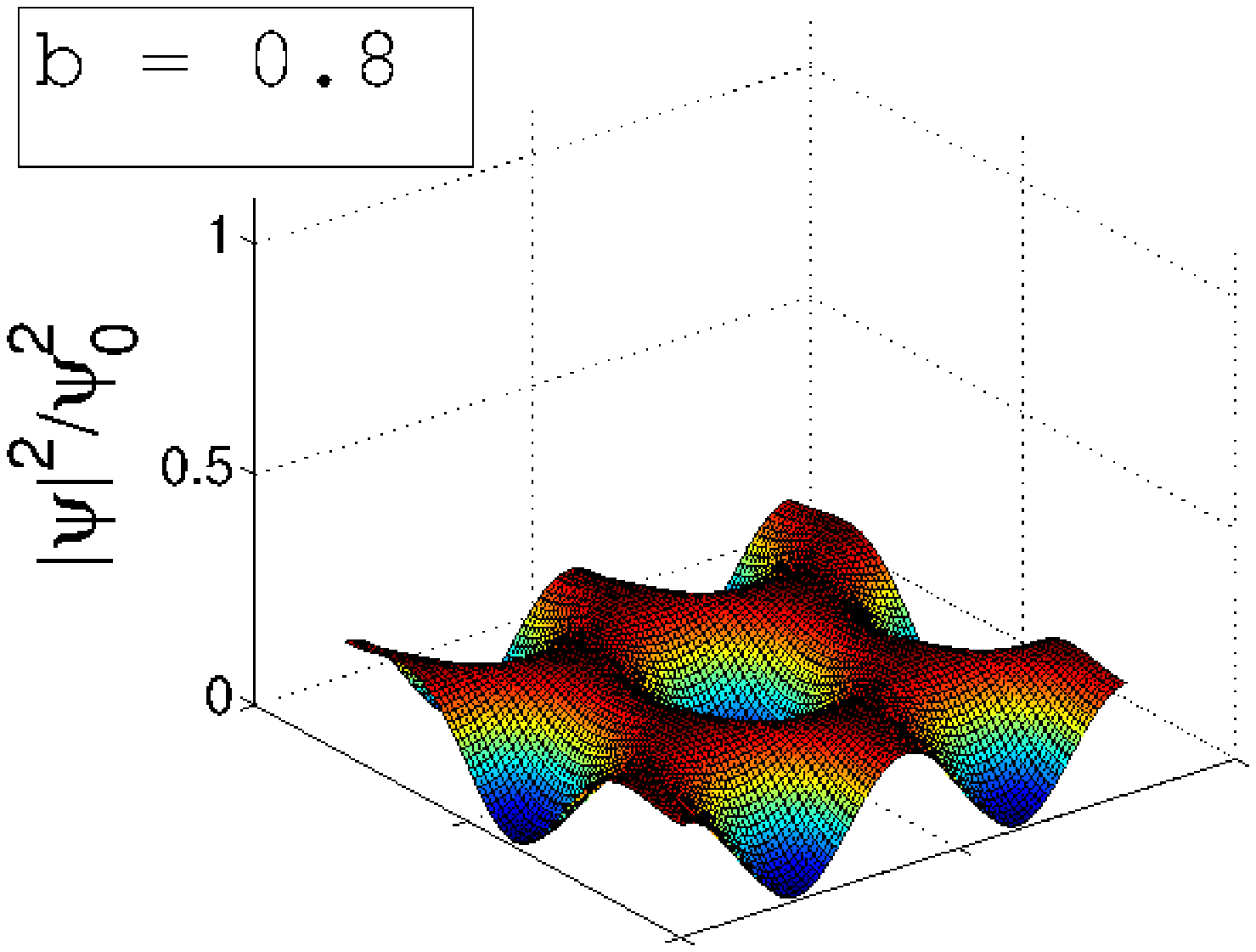}
\caption{(Color online) Spatial variation of the normalized superfluid
density $|\psi(\mathbf{r})|^2/\psi_0^2$ in a hexagonal FLL at four different
reduced fields $b = \langle B\rangle / B_{c2}$. The minima of
$|\psi(\mathbf{r})|^2$ correspond to positions of vortex cores.}
\label{fig:OrdParam}
\end{figure}

For the limiting cases of $\kappa \rightarrow \infty$ ($\kappa = \lambda/\xi$
is the GL parameter) and $B_{c1} < B \ll B_{c2}$ simplified, second moment
analysis methods were developed.\cite{Brandt88,Brandt03} In the general case
of arbitrary $\kappa$ and $B$, the solution is more complicated and various
approximations have been suggested.\cite{BrandtJLTPhys88and77,
Sidorenco90,HaoClem91,Reotier97review,Yaouanc97,Sonier00RMP} A feasible and
precise minimization algorithm of the ``classical'' GL free energy has been
suggested by Brandt.\cite{Brandt97_NGLmethod} The method was first used in the
experimental work in Ref.\ \onlinecite{Laulajainen06}. The difference between
the SC and the normal-state free energies $\Delta F = F_s - F_n$ is expressed
as (in SI units):\cite{deGennes,Tinkham,Brandt97_NGLmethod}

\begin{equation}\label{eq:GLfunctional}
\Delta F = \alpha |\psi|^2+\frac{\beta}{2} |\psi|^4 +
\frac{1}{2m^*}\left|\left(\frac{\hbar}{i}\nabla-2e\mathbf{A}\right)
\psi\right|^2+\frac{\mathbf{B}^2}{2\mu_0}.
\end{equation}

\begin{figure}[tb]
\includegraphics[width=0.8\linewidth]{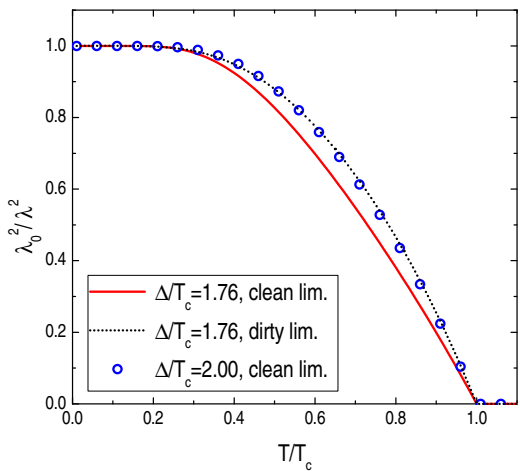}
\caption{(Color online) Temperature dependence of normalized superfluid
density $\lambda_0^2/\lambda^2(t)$ in clean and dirty limits ($t = T/T_c$). In
open circles $\lambda_0^2/\lambda^2(t)$ in the clean limit but larger
gap-to-$T_c$ ratio is shown.}
\label{fig:sfldDirty1}
\end{figure}

Here, $\mathbf{B} = \text{rot}\mathbf{A}$, the parameter $\beta =
\mu_0/2\cdot(\kappa e\hbar/m)^2$ is determined by the GL parameter $\kappa$,
and $\psi_0^2 = -\alpha/\beta>0$. The parameter $\psi_0^2$ is the superfluid
density deep in the bulk of the superconductor in the limit of low fields
(i.e.\ in the Meissner state), which is related to the magnetic penetration
depth $\lambda$. The relation between $\lambda^{-2}$ and $\psi_0^2$ is (in SI
units):\cite{deGennes}

\begin{equation}
\lambda^{-2} = \frac{4 \mu_0 e^2}{m^*}\psi_0^2.
\end{equation}

In an applied field, however, the superfluid density $|\psi|^2$ is spatially
inhomogeneous with minima at the vortex cores due to the formation of a
FLL.\cite{Abrikosov} Fig.\ \ref{fig:OrdParam} shows the spatial variation of
$|\psi(\mathbf{r})|^2/\psi_0^2$ at different reduced fields $b = \langle
B\rangle/B_{c2}$ in the limit of $\kappa \rightarrow \infty$ corresponding to
the minimum of Eq.\ (\ref{eq:GLfunctional}) ($\langle B\rangle$ is the mean
field in the sample). Although $\lambda$ is field-independent (as well as
$\alpha$ and $\beta$) and finite at $T = T^{B_{c2}}_c \equiv T_c (B \neq 0)$,
the superfluid density reduces with increasing field and vanishes at $T =
T^{B_{c2}}_c$. Therefore, with increasing field for $b\gtrsim 0.05$ (e.g.\
for a non-high-$T_c$ superconductor; see Fig.\ \ref{fig:OrdParam}) the
correction factor $(1-b)$ to the superfluid density becomes significant. The
mean value of the superfluid density reduces with increasing field as
follows:\cite{Brandt97_NGLmethod,Brandt03}

\begin{equation}
\rho_s = \langle |\psi|^2 \rangle \simeq (1-b)\psi_0^2.
\end{equation}

For small values of $b \rightarrow 0$ and high $\kappa$, as in most of the
high-$T_c$ superconductors, we have $\rho_s \propto \lambda^{-2}$. In the
present analysis the free energy [Eq.\ (\ref{eq:GLfunctional})] for the given
$\lambda$, $\xi$, and $\langle B\rangle$ was minimized using the method
suggested by Brandt.\cite{Brandt97_NGLmethod} This results in a solution for
spatial variation of the field $\mathbf{B}(\mathbf{r})$ and the order
parameter $\psi(\mathbf{r})$.

\section{Some details on Eq.\ (\ref{eq:SwaveSfld})}

We use Eq.\ (\ref{eq:SwaveSfld}) suggested in Ref.\ \onlinecite{Xu95} for the
case of arbitrary scattering rate $1/\tau$ (mean free path $l = v_F \tau$).
For the classical BCS gap-to-$T_c$ ratio $\Delta_0/k_BT_c = 1.76$ the
temperature dependence of the normalized superfluid density
$\lambda_0^2/\lambda^2(t)$ obtained with Eq.\ (\ref{eq:SwaveSfld}) in clean
($\tau \gg \xi_\text{cl}/v_F$) and dirty ($\tau \ll \xi_\text{cl}/v_F$) limit
is given in Fig.\ \ref{fig:sfldDirty1} ($t=T/T_c$). The results are in good
agreement with curves given in Ref.\ \onlinecite{Tinkham}. Note, the shape of
$\lambda_0^2/\lambda^2(t)$ depends on the scattering rate only for $\tau \sim
\xi_\text{cl}/v_F$. For the current precision of measurement the parameters
$\Delta_0$ and $\tau$ are correlated. The dirty-limit curve with
$\Delta_0/k_BT_c=1.76$ can be well fitted with the clean-limit model with
$\Delta_0/k_BT_c = 2.0$ (see Fig.\ \ref{fig:sfldDirty1}).



\begin{thebibliography}{99}
\bibitem{Footnote1}
Small and unphysical reduction ($\simeq 5$\,{\%} in the field range of 50 to
640\,mT) of $\lambda$ with increasing field was obtained for the value of
$\xi$ determined from the corresponding $B_{c2}$ (using the GL relation
$B_{c2} = \Phi_0/2\pi\xi^2$). Such deviation rather points to a slight
systematic deviation of vortex lattice field distribution from that of the
ideal GL vortex lattice.
\bibitem{Remark-on-n}
For the charge carrier concentration we assume $n$ = 4 electron holes
per cell ($Z$ = 2) for BaPt$_4$Ge$_{12}$ and a decrease of $n$ to only 2
when one Pt atom is substituted by Au.
\end{thebibliography}
\end{document}